\input{aipcheck}

\documentclass[
    ,final            
  ]
  {aipproc}

\layoutstyle{6x9}

\begin{document}

\title{Transverse Single Spin Asymmetry Measurement with J/$\Psi$ in Polarized p+p Collisions at RHIC}

\classification{14.20.Dh} \keywords {Proton structure, transverse
spin}

\author{Han Liu for the PHENIX Collaboration}{
  address={New Mexico State University, Las Cruces, NM 88003, USA}
}

\begin{abstract}
The PHENIX experiment has measured transverse single spin
asymmetry of J/$\Psi$ in polarized p+p collisions at forward
rapidity at $\sqrt{s}=200$ GeV. The data were collected from year
2006 run of RHIC with average beam polarization of 56\%. At RHIC
energy, heavy quark production is dominated by gluon gluon
interaction. Therefore, the transverse single spin asymmetry in
J/$\Psi$ production can provide a clean measurement of the gluon
Sivers distribution function.
\end{abstract}

\maketitle


\section{INTRODUCTION}

The measurement of transverse single spin asymmetries gives us an
opportunity to probe the quark and gluon structure of transversely
polarized nucleons. Large transverse single spin asymmetries of up
to 20\% - 40\% were observed for pions produced at large $x_F$ at
$\sqrt{s}$ = 20 GeV\cite{E704} and have been found to persist at
$\sqrt{s}$ = 200 GeV by the STAR \cite{Star} and
BRAHMS\cite{BRAHMS} experiments, although they were expected to
vanish in pQCD calculations at the leading order. A number of pQCD
based models have been developed to explain this phenomenon. Among
them are the Sivers effect (transversely asymmetric $k_T$ quark
and gluon distributions)\cite{Sivers}, the Collins effect
(transversity distribution in combination with spin-dependent
fragmentation function)\cite{Collins}, and the higher twist effect
(interference between quark and gluon fields in the initial or
final state)\cite{Twist1,Twist2}.

At RHIC energy, heavy flavor production is dominated by
gluon-gluon interaction, thus Collins effect has minimum impact on
$A_N$ as the gluon's transversity is zero. Therefore, the
production of heavy flavor particles in transversely polarized p+p
collisions at the PHENIX experiment offers a good opportunity to
probe the gluon's Sivers effect.

\section{MEASUREMENT AND ASYMMETRY CALCULATION}

In this report we present the latest results of transverse single
spin asymmetry in J/$\Psi$ production in polarized p+p collisions
at RHIC. The PHENIX muon spectrometers have measured J/$\Psi$
yields through the J/$\Psi\rightarrow \mu^+\mu^-$ channel at
forward rapidities ($1.2<|\eta|<2.4$). During the recent 2006
polarized pp run, the PHENIX experiment has collected 2.7
pb$^{-1}$ data with transverse beam polarization of about 56\%.
The Level-2 filterd data sample has been used in this analysis.

The J/$\Psi$ yield is obtained
by fitting the dimuon mass spectrum with a single exponential
background plus two gaussian functions (for J/$\Psi$ and
$\Psi^{\prime}$) in the mass range 2.3 GeV - 4.5 GeV (as shown in
figure 1). The total Number of J/$\Psi$ from this fit is 5236
$\pm$ 81. The Monto Carlo simulation shows the systematic error in
determining J/$\Psi$ yield is less than 2\%.
\begin{figure}[tb]
  \includegraphics[height=.3\textheight]{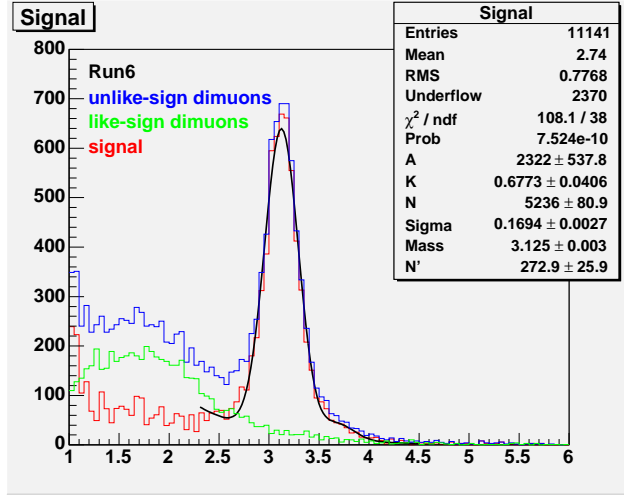}
  \caption{Dimuon invariant mass distribution.}
  \label{mass}
\end{figure}


 The transverse single spin asymmetry $A_N$ can
be determined by:
\begin{equation}
A_N= {1\over P_b}{{\sigma^\uparrow-\sigma^\downarrow}
\over{\sigma^\uparrow+\sigma^\downarrow}} = {1\over
P_b}{{N^\uparrow-R N^\downarrow} \over{N^\uparrow+R
N^\downarrow}}, \label{lum}
\end{equation}
where $P_b$ is the beam polarization,
$\sigma^\uparrow(\sigma^\downarrow)$ is the production cross
section, $N^\uparrow(N^\downarrow)$ is the J/$\Psi$ yield from
up(down) polarized bunches, and $R=L^\uparrow / L^\downarrow$ is
the relative luminosity of bunch crossings of opposite
polarization sign. The relative luminosity is measured by two
global detectors, BBC and ZDC.

Alternatively, we also use square root formula
\begin{equation}
 A_N={1\over P_b}{{\sqrt{N_L^\uparrow\cdot
N_R^\downarrow}-\sqrt{N_L^\downarrow\cdot N_R^\uparrow}}
\over{\sqrt{N_L^\uparrow\cdot
N_R^\downarrow}+\sqrt{N_L^\downarrow\cdot N_R^\uparrow}}},
\label{sqrt}
\end{equation}
to cross check our results.

The asymmetry of J/$\Psi$ was determined for each fill using Eq.
(\ref{lum}), then averaged over all fills.

The raw J/$\Psi$ sample is contaminated by other background,
mainly comes from Drell-Yan process, open charm and light hadrons
decay. The $A_N^{J/\Psi}$ was corrected for the contribution of
background by using
\begin{equation}
\label{bg-1} A_N^{J/\Psi}={{A_N^{incl}-r\cdot
A_N^{BG}}\over{1-r}},
\end{equation}
\begin{equation}
\label{bg-2} \delta A_N^{J/\Psi}={{\sqrt{(\delta
A_N^{incl})^2+r^2\cdot (\delta A_N^{BG})^2}}\over{1-r}},
\end{equation}
where $r$ is the background fraction under the J/$\Psi$ peak,
$A_N^{BG}$ is the averaged background asymmetry. $A_N^{BG}$ is
obtained using the like (unlike) sign dimuon pairs within 2.0 GeV$
< M_{\mu\mu} < 2.5$ GeV,
 and like sign dimuon pairs within 2.8 GeV$ < M_{\mu\mu} < 3.5$
GeV.

The systematic uncertainty has been checked with bunch shuffling
technique with randomly assigned polarization direction for
bunches in each fill. The bunch-to-bunch and fill-to-fill
systematic errors are much smaller than the statistical errors for
this data set.

\section{RESULTS}

\begin{figure}[h]
  \includegraphics[height=.3\textheight]{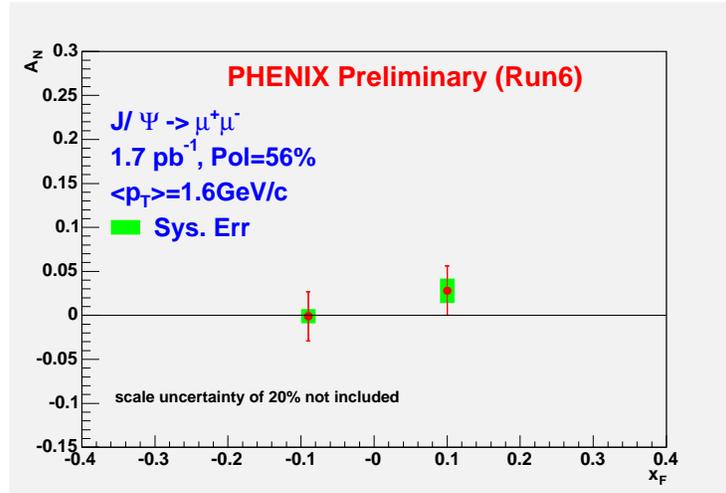}
  \caption{Transverse single spin asymmetry vs. $x_F$ for J/$\Psi$ at forward rapidities.
  A scale uncertainty of 20\% is not included.}
  \label{fig1}
\end{figure}

The asymmetry was calculated for two $x_F$ bins. Figure \ref{fig1}
shows the preliminary result of the single spin asymmetry as a
function of $x_F$ in J/$\Psi$ production from the 2006 RHIC
transverse run. A scale uncertainty of 20\% due to the absolute
polarization values is not included in this analysis. Within the
limits of errors, $A_N$ is consistent with zero over the measured
$x_F$ range. \vskip 0.3cm

\vskip 0.3cm

\begin{figure}[h]
  \includegraphics[height=.3\textheight]{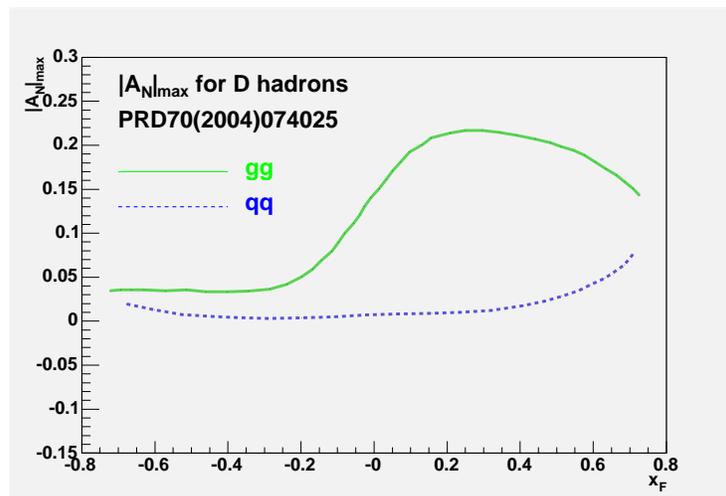}
  \caption{Maximized values of $|A_N|$ for the process p+p$\rightarrow$DX as a function of $x_F$ at fixed transverse momentum $p_T=1.5$ GeV/c
  at RHIC energy. }
  \label{fig2}
\end{figure}

 Currently there is no theoretical calculation of
J/$\Psi$'s $A_N$. However, there is a prediction of maximum values
of $|A_N|$ in D meson production at fixed transverse momentum
$p_T=1.5$ GeV/c at RHIC energy, as shown in figure
\ref{fig2}\cite{Anselmino}. The solid line shows $|A_N|_{max}$
when the gluon Sivers function is set to its maximum with the
quark Sivers function set to zero, and the dashed line corresponds
to a maximized quark Sivers function with the gluon Sivers
function set to zero.  If the transverse single spin asymmetry
comes from the initial state (the J/$\Psi$ production mechanism
does not play an important role), then we expect $A_N$ in J/$\Psi$
and D meson production to be similar. The fact that there is no
sizable asymmetry observed in figure \ref{fig1} may indicate our
results disfavor the maximum contribution of gluon's Sivers
function.

\section{SUMMARY}
The transverse single spin asymmetry $A_N$ for J/$\Psi$ has been
measured for the first time at forward rapidity and $\sqrt{s}$ =
200 GeV with the PHENIX muon spectrometers. Using Level-2 dimuon
triggered data sample (corresponds to 60\% of the whole 2006
transverse data set), no sizable $A_N$ of J/$\Psi$ has been
observed
which indicates our results disfavor the maximum gluon Sivers
contribution.



\bibliographystyle{aipproc}   

\bibliography{sample}

\IfFileExists{\jobname.bbl}{}
 {\typeout{}
  \typeout{******************************************}
  \typeout{** Please run "bibtex \jobname" to optain}
  \typeout{** the bibliography and then re-run LaTeX}
  \typeout{** twice to fix the references!}
  \typeout{******************************************}
  \typeout{}
 }

\end{document}